\documentclass[floatfix,aps,amsmath,nofootinbib,twocolumn,10pt]{revtex4}

\usepackage{listings}
\usepackage{graphicx}
\usepackage{bm}
\usepackage{rotating}
\usepackage{array}
\usepackage{amsmath}
\usepackage{amssymb} %花体字母加粗
\usepackage{mathrsfs} %花体字母
\usepackage{cancel}
\usepackage{subfig}
\usepackage{float}
\usepackage{caption}
\usepackage[dvipsnames]{xcolor}
\usepackage[normalem]{ulem}

\def\({\left(}
\def\){\right)}
\def\[{\left[}
\def\]{\right]}

\def\e{\begin{equation}}
\def\q{\end{equation}}
\def\m{\begin{eqnarray}}
\def\n{\end{eqnarray}}

\begin{document}
\title{Diversity of Fuzzy Dark Matter Solitons}% Force line breaks with \\
\author{Chen Tan$^{1,2,3,4}$}
\author{M. Le Delliou$^{1,2,3,4,5,6}$}
%\email{delliou@lzu.edu.cn,Morgan.LeDelliou.IFT@gmail.com}
\author{Ke Wang$^{7}$}
\thanks{Corresponding author: {wangke@lnnu.edu.cn}}
\affiliation{$^1$School of Physical Science and Technology, Lanzhou University, Lanzhou 730000, China}
\affiliation{$^2$Institute of Theoretical Physics $\&$ Research Center of Gravitation, Lanzhou University, Lanzhou 730000, China}
\affiliation{$^3$Key Laboratory of Quantum Theory and Applications of MoE, Lanzhou University, Lanzhou 730000, China}
\affiliation{$^4$Lanzhou Center for Theoretical Physics $\&$ Key Laboratory of Theoretical Physics of Gansu Province, Lanzhou University, Lanzhou 730000, China}
\affiliation{$^5$Instituto de Astrof\'isica e Ci\^encias do Espa\c co, Universidade de Lisboa,
Faculdade de Ci\^encias, Ed.~C8, Campo Grande, 1769-016 Lisboa, Portugal}
\affiliation{$^6$Universit\'e de Paris-Cit\'e, APC-Astroparticule et Cosmologie (UMR-CNRS 7164), 
%Batiment Condorcet, 10 rue Alice Domon et L\'eonie Duquet, F-75205 Paris Cedex 13, France.
F-75006 Paris, France}
\affiliation{$^7$Department of Physics, Liaoning Normal University, Dalian 116029, China}

\date{\today}% It is always \today, today,

\begin{abstract}
According to the Schrödinger-Poisson equations, fuzzy dark matter (FDM) can form a stable equilibrium configuration, the so-called FDM soliton. In principle, given the FDM particle mass, the profile of the FDM soliton is fixed. In practice, however, there is a great diversity of structures in the Universe. Possible %In this paper, we enumerate some possible 
causes of such diversity can lie in %, 
such %as the 
%effects 
sources as %of 
the gravitoelectric field due to a central supermassive black hole, the gravitomagnetic field due to the system angular momentum, an extra denser and compact FDM soliton and an ellipsoidal baryon background. We %And we 
find that the effects of the gravitomagnetic field due to the soliton's self-angular momentum are very weak while those %but the effects 
of the other source%cause
s are considerable.
\end{abstract} 

\maketitle

%\tableofcontents

\section{Introduction}
\label{sec:intro}
The existence of dark matter (DM) is strongly suggested %can be inferred 
from %the 
rotation curves of galaxies~\cite{Rubin:1982kyu}, %the 
evolution of large scale structure~\cite{Davis:1985rj} and %the 
gravitational lensing observations~\cite{Clowe:2006eq}. Depending on %According to 
the free streaming length of DM, it %DM 
can be divided into cold, warm and hot categories. Although the standard Lambda cold DM ($\Lambda$CDM) cosmological model is very successful and the latest cosmic microwave background (CMB) observations~\cite{Planck:2018vyg} suggests that CDM accounts for about $26\%$ of today's energy density in the Universe, we still have no certainty on the existence or nature of CDM, whether sourced by particles, other objects or explained away by gravity modifications \cite{Sanders:2006sz}. %don't know what particle or object CDM is. 
Especially, as one of the most promising candidates for CDM particle, the weakly interacting massive particles (WIMPs) grounded on supersymmetric theories of particle physics still have not been detected~\cite{PandaX-II:2016vec,LUX:2015abn,LUX:2016ggv}. Moreover, primordial black holes (BHs) can also serve as CDM~\cite{Carr:2016drx}, although % but 
they still have not been identified. 
These null results coupled with the failures of CDM particles on sub-galactic scales~\cite{Primack:2009jr,Bull:2015stt} imply that DM may not entirely fall under the CDM paradigm%be cold
.

A %There is a 
promising alternative to CDM lies in %which is 
an ultralight scalar field with spin-$0$, %\Mov{with }
extraordinarily light mass ($\sim10^{-22}\rm{eV}/c^2$) and de Broglie wavelength comparable to a few kpc, coined %namely 
fuzzy DM (FDM)~\cite{Hu:2000ke}.
Due to its %the 
large occupation numbers in galactic halos, FDM behaves as a classical field
obeying the coupled Schrödinger–Poisson (SP) system of equations,
\begin{equation}
\label{eq:sp00} 
\begin{cases}
\begin{aligned}
&i\hbar\frac{\partial \Psi}{\partial t}=\left(-\frac{\hbar^2}{2m}\nabla^2+m\Phi\right)\Psi , \\
&\nabla^2 \Phi=4\pi G |\Psi|^2, 
\end{aligned}
\end{cases}
\end{equation}
where $m$ is the mass of the FDM particle, which is described by the wavefunction $\Psi$, and the gravitational
potential $\Phi$ is sourced by the FDM density $\rho=|\Psi|^2$. According to Eq.~\eqref{eq:sp00}, FDM can condensate into a many particles ground state coined as soliton.
The SP system can also describe %can govern 
the evolution of FDM in an expanding Universe~\cite{Schive:2014dra,Schive:2014hza,DeMartino:2017qsa,Mocz:2019pyf}.
Note %It is worth noting 
that the SP system is just the weak field limit of the general relativistic Einstein–Klein–Gordon (EKG) system~\cite{Kaup:1968zz,Ruffini:1969qy,Ma:2023vfa}\,  and that
the SP system %\Mov{being }%
should be 
replaced by the Gross-Pitaevskii-Poisson (GPP) system when FDM particles %the 
self-interactions %between FDM particles 
exist~\cite{Chavanis:2011zi,Chavanis:2011zm}.

In this paper, for %For 
simplicity, %in this parer, 
we confine ourself to the SP system in a non-expanding Universe. The evolution of this system is usually simulated numerically, including the formation, %the 
perturbation, %the 
interference/collision and %the 
tidal disruption/deformation of FDM solitons~\cite{Guzman:2004wj,Paredes:2015wga,Edwards:2018ccc,Munive-Villa:2022nsr}. Focussing on %If one just care about 
stationary solutions, one can turn to the shooting method to compute %find 
the eigenvalues of equilibrium configurations~\cite{Guzman:2004wj,Davies:2019wgi}. 
In this work we restrict to different situations where numerical treatment can be avoided cleverly, as illustrated below.

Since FDM solitons %are 
only depend %ent 
on the mass of FDM, the SP system obeys %has the 
scaling symmetry. As a result, the Universe should {\it a priori} present %be 
very monotonous and %full of 
similar structures.
Yet, %But 1in fact\Mov{,} 
there is a great diversity of structures in the Universe. Therefore, although we do not here consider the modifications of gravity or various possible FDM interactions, the real astrophysics systems must follow %the different 
variants of the exact SP system%, even though the modified gravity and the different interactions between FDM are not considered
. In this paper, we will consider several variants, induced by % due to different 
extra terms, such as the gravitoelectric field, the gravitomagnetic field, both sourced by the Weyl tensor \cite{Ellis2012}, an extra FDM soliton and the baryon density profile.
Our motivations are as follows.
First since, as pointed out in~\cite{Guzman:2004wj,Salehian:2021khb}, the SP system is the weak field limit of the EKG system%. Therefore
, there must exist %be 
a variant system, between these two systems, for which %ends where the 
gravitoelectromagnetism is a good approximation. While the gravitoelectric field or the gravitational potential of a supermassive BH has been discussed~\cite{Edwards:2018ccc,Munive-Villa:2022nsr,Davies:2019wgi}, the gravitomagnetic field due to the system's angular momentum has not yet been considered% before
. Secondly, although the interference/collision with an extra FDM soliton has also %has 
been studied~\cite{Paredes:2015wga,Edwards:2018ccc,Munive-Villa:2022nsr}, those studies restricted to %they just dealt with 
the cases in which %where 
the density-ratio of two solitons is $\sim1$. For the extreme cases where the density-ratio of the denser solitons to another one is $\gtrsim10^4$, numerical simulations would require a large enough %need a larger 
simulation box to contain the larger soliton with lower density, while requiring % or need 
a higher resolution to depict the smaller soliton with higher density. However, a simulation combining a large %with a larger 
simulation box and high %er 
resolution at the same time is prohibitively expensive and almost impossible to solve. %In this work we therefore restrict to situations where numerical treatment can be avoided cleverly}. 
Finally, when %a 
non-spherical baryon profile is located in galaxies, it % which 
results in non-spherical FDM solitons. The %, for example the 
impact of a baryon profile with %the 
cylindrical symmetry has been quantified in~\cite{Bar:2019bqz}. 
In fact, the baryon profile in the Milky Way may be ellipsoidal~\cite{Iocco:2015xga,Lin:2019yux}.

This paper is organized as follows.
In section~\ref{sec:ge}, we briefly review the effects of the gravitoelectric field due to a supermassive BH on the FDM solitons.
In section~\ref{sec:gm}, we %prove
study the effects of the gravitomagnetic field due to the rotation velocity of FDM, the supermassive BH spin and the orbital motion of supermassive
BH binary on the FDM solitons.
In section~\ref{sec:edr}, we study the interaction between soliton binary with the extreme density-ratio.
In section~\ref{sec:bar}, we calculate the FDM solitons in a given ellipsoidal baryon profile. 
Finally, a brief summary and discussions are included in section~\ref{sec:sd}.

\section{Effects of the gravitoelectric field}
\label{sec:ge}
The effects of the gravitoelectric field include the tidal disruption/deformation of FDM solitons by a nearby supermassive BH~\cite{Edwards:2018ccc,Munive-Villa:2022nsr} and the modified formation of FDM solitons by a central supermassive BH~\cite{Davies:2019wgi}. While the former non-spherical impacts are more complicated and have to be studied by numerical simulations, the latter spherical ones are simple and can be studied by the shooting method. In this section, we briefly review the latter cases.

\begin{figure*}[]
\begin{center}
\subfloat{\includegraphics[width=0.6\textwidth]{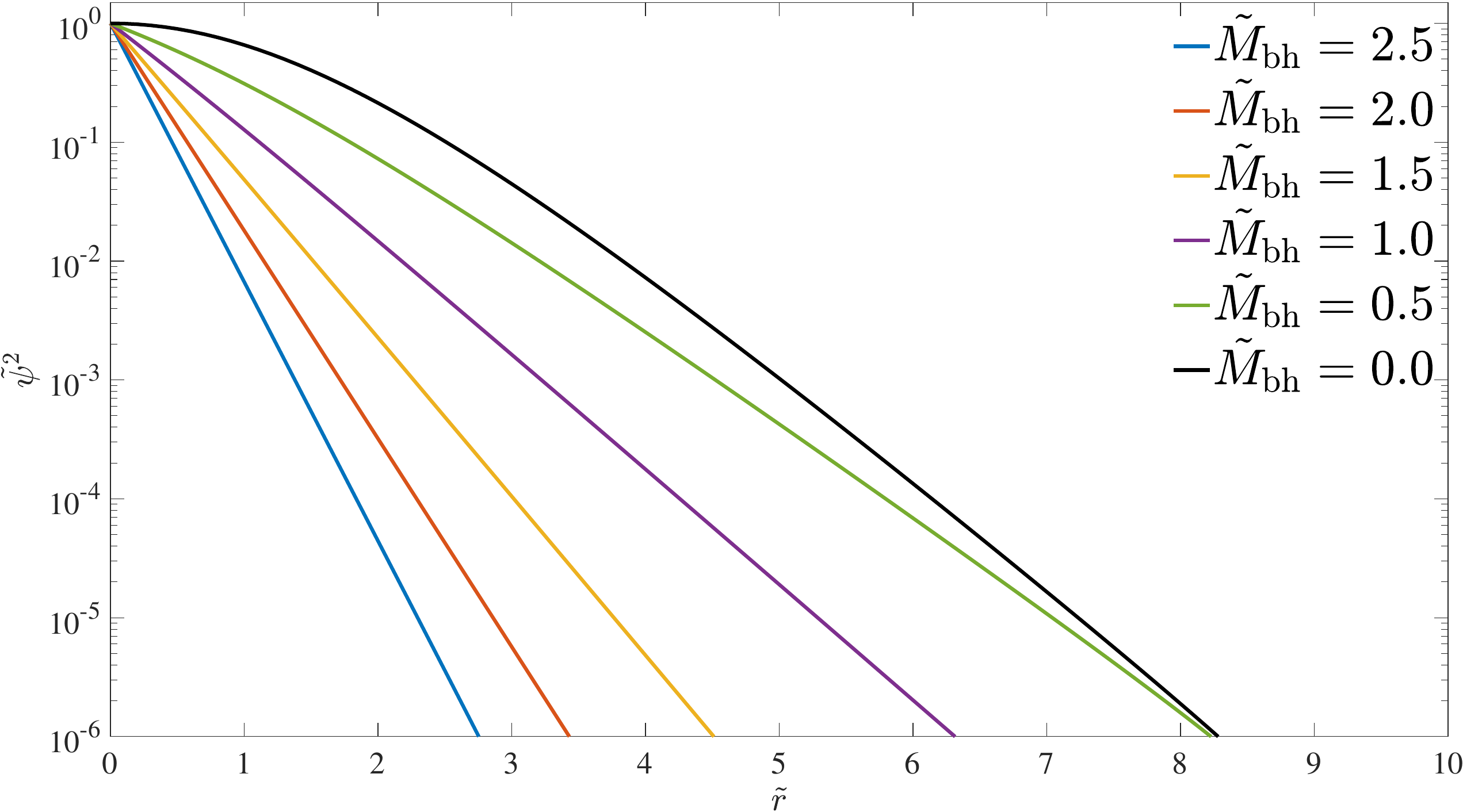}}
\end{center}
\captionsetup{justification=raggedright}
\caption{Density %The density 
profiles of the ground state solutions with different dimensionless supermassive BH mass $\tilde{M}_{\rm bh}$.}
\label{fig:ge}
\end{figure*}

Firstly, the gravitoelectric field $\mathbf{E}_{\rm g}$ derived from %or 
the gravitational potential $\Phi_{\rm e}$, sourced by a central supermassive BH, contributes to a variant of the exact SP system as 
\begin{equation}
\label{eq:sp10} 
\begin{cases}
\begin{aligned}
&i\hbar\frac{\partial \Psi}{\partial t}=\left[-\frac{\hbar^2}{2m}\nabla^2+m(\Phi+\Phi_{\rm e})\right]\Psi,\\
&\nabla^2 \Phi=4\pi G |\Psi|^2,
\end{aligned}
\end{cases}
\end{equation}
where the supermassive BH with mass $M_{\rm bh}$ has a point mass potential
\begin{equation}
\label{eq:P10}
\Phi_{\rm e}=-\frac{GM_{\rm bh}}{r}.
\end{equation}
When the system in question features %the 
spherical symmetry, the ansatz of $\Psi(r, t) =e^{-i\gamma t/\hbar}\psi(r)$, where $\gamma$ can be interpreted as the ansatz energy eigenvalue, leads to %means 
the FDM soliton density %is 
$\rho(r)=|\Psi|^2=\psi^2(r)$ and the FDM soliton mass %is 
$M=\int_0^\infty4\pi r^2\rho(r)dr$. After defining the system's %a number of 
dimensionless variables as
\begin{align}
&\tilde{\psi}\equiv\frac{\hbar\sqrt{4 \pi G}}{mc^2} \psi,\label{eq:dimLessPsi}\\
&\tilde{r}\equiv\frac{mc}{\hbar}r,\\
&\tilde{\Phi}\equiv\frac{1}{c^2}\Phi,\\
&\tilde{\gamma}\equiv\frac{1}{mc^2}\gamma,\\
&\tilde{M} \equiv \frac{GMm}{\hbar c}\label{eq:dimLessM},\\
&\tilde{M}_{\rm bh} \equiv \frac{GM_{\rm bh}m}{\hbar c},
\end{align}
the dimensionless version of Eq.~(\ref{eq:sp10}), with the spherical ansatz, follows% is
\begin{equation}
\label{eq:sp11} 
\begin{cases}
\begin{aligned}
&\frac{\partial^2 (\tilde{r}\tilde{\psi})}{\partial \tilde{r}^2}=2\tilde{r}\left(\tilde{\Phi}-\tilde{\gamma}-\frac{\tilde{M}_{\rm bh}}{\tilde{r}}\right)\tilde{\psi},\\
&\frac{\partial^2 (\tilde{r}\tilde{\Phi})}{\partial \tilde{r}^2}=\tilde{r}\tilde{\psi}^2.
\end{aligned}
\end{cases}
\end{equation}
Fulfilling the arbitrary normalization $\tilde{\psi}(\tilde{r}=0)=1$ and
$\tilde{\Phi}( \tilde{r} = 0) = 0$\footnote{\label{ftn:norm}This normalization condition for the soliton self-potential, $\Phi$, is different from the conventional choice of normalization as for the BH potential, $\Phi_{\rm e}$, that $\Phi_{\rm e}(r=\infty)=0$. Although it produces a nonvanishing asymptotic value at infinity, $\tilde{\Phi}( \tilde{r} = \infty)$, this difference does not affect the solutions with the normalization $\tilde{\psi}(\tilde{r}=0)=1$.}, the boundary conditions $\tilde{\psi}( \tilde{r} = \infty) = 0$,
$\frac{\partial \tilde{\psi}}{\partial \tilde{r}}|_{\tilde{r} = 0} = 0$ and $\frac{\partial \tilde{\Phi}}{\partial \tilde{r}}|_{\tilde{r} = 0} = 0$, choosing the supermassive BH mass $\tilde{M}_{\rm bh}=\{0.0,0.5,1.0,1.5,2.0,2,5\}$ as the input parameter and adjusting the quantized eigenvalue $\tilde{\gamma}$, we can calculate the equilibrium configurations from Eq.~(\ref{eq:sp11}) by the shooting method. Given the input parameter $\tilde{M}_{\rm bh}$, we obtain the only stable solution from %only the solution from 
the smallest $\tilde{\gamma}$, which is %is stable and 
the ground state. In Tab.~\ref{tb:ge}, we list the input values of $\tilde{M}_{\rm bh}$ and the corresponding eigenvalue $\tilde{\gamma}$ and soliton mass $\tilde{M}$ of the ground state solutions. In Fig.~\ref{fig:ge}, the corresponding soliton profiles are plotted. We find that the larger $\tilde{M}_{\rm bh}$ results in a %leads to a more 
denser and more compact soliton, leading to %hence 
a smaller dimensionless soliton mass $\tilde{M}$, due to the fixed normalization $\tilde{\psi}(\tilde{r}=0)=1$.
The solutions under this normalization can be rescaled by the following scaling symmetry %verified by
%Since 
%this variant SP system (Eq.~\ref{eq:sp11}) %\Mov{obeys }%has 
%the scaling symmetry
\begin{align}
&\tilde{\psi}  \longrightarrow  \lambda \tilde{\psi}\label{eq:PsiTildeScaling}, \\
&\tilde{r}  \longrightarrow  \lambda^{-1/2} \tilde{r},\\
&\tilde{\Phi}  \longrightarrow  \lambda \tilde{\Phi}, \\
&\tilde{\gamma}  \longrightarrow  \lambda \tilde{\gamma}, \\
&\tilde{M} \longrightarrow  \lambda^{1/2} \tilde{M}\label{eq:MtildeScaling},\\
&\tilde{M}_{\rm bh} \longrightarrow  \lambda^{1/2} \tilde{M}_{\rm bh}.
\end{align}%
Taking into account a simultaneously increasing $\lambda$,
%In fact, 
%\Mov{However, }%As for 
%its
the physical mass $M$ actually %, in fact it 
increases with $\tilde{M}_{\rm bh}$~\cite{Davies:2019wgi}.

%which allows to bring the system its values for the real central density through Eqs.~\eqref{eq:dimLessPsi} and \eqref{eq:PsiTildeScaling}. The corresponding scaling must then increase to trace the density increase induced by the BH, resulting in the physical mass increase through Eqs.~\eqref{eq:dimLessM} and \eqref{eq:MtildeScaling}.}
\begin{table}[htbp] 
\renewcommand\arraystretch{1.5}
\captionsetup{justification=raggedright}
\caption{Input values of the dimensionless supermassive BH mass $\tilde{M}_{\rm bh}$ and the corresponding eigenvalue $\tilde{\gamma}$ and soliton dimensionless mass $\tilde{M}$ of the ground state solutions.} 
\label{tb:ge}
\begin{tabular}{ | m{1cm} | m{1.5cm}| m{1.5cm} | } 
\hline $\tilde{M}_{\rm bh}$ & $\tilde{\gamma}$ & $\tilde{M}$\\ 
\hline 
0.0 & \ 0.6495 & 2.0622\\ 
0.5 & \ 0.1164 & 0.7916\\ 
1.0 & -0.5113 & 0.2169\\ 
1.5 & -1.0839 & 0.0718\\ 
2.0 & -1.9767 & 0.0309\\ 
2.5 & -3.1101 & 0.0159\\ 
\hline
\end{tabular}
\end{table}

\section{Effects of the gravitomagnetic field}
\label{sec:gm}
The SP system corresponds to %is 
the weak field limit of the general relativistic EKG system~\cite{Guzman:2004wj}. 
A %Therefore, there must be a 
variant system can therefore account for %between these two extreme ends by taking 
some general relativistic corrections to the SP system, on the spectrum between the full EKG system and its weak field approximation, such as %into consideration, for example 
the gravitomagnetic field $\mathbf{B}_{\rm g}$.
Such %Then there is a 
variant of the exact SP system reads%as
\begin{equation}
\label{eq:sp20} 
\begin{cases}
\begin{aligned}
&i\hbar\frac{\partial \Psi}{\partial t}=\left[-\frac{\hbar^2}{2m}\nabla^2+m(\Phi+\Phi_{\rm e}+\Phi_{\rm m})\right]\Psi ,\\
&\nabla^2 \Phi=4\pi G |\Psi|^2,
\end{aligned}
\end{cases}
\end{equation}
where the gravitational potential $\Phi_{\rm m}$ is related to the gravitomagnetic field $\mathbf{B}_{\rm g}$ and the angular momentum\footnote{Note that $\Psi=\sqrt{N}\Psi_1$ \cite{Ruffini:1969qy}, where $\Psi_1$ is the dimensional wavefunction of one FDM particle and $N$ is the fixed number of particles in the soliton. As $N$ is constant, we have $i\hbar\frac{\partial \Psi}{\partial t}=i\hbar\frac{\partial \sqrt{N}\Psi_1}{\partial t}=  \sqrt{N}i\hbar\frac{\partial \Psi_1}{\partial t}$. Therefore the Schr\"odinger Eq.~\eqref{eq:sp20} can reduce to the description of a single particle of the soliton and the $\hat{\mathbf{L}}$ operator brings out the angular momentum for a single particle.} of an FDM particle 
$\hat{\mathbf{L}}$.  In the case of a non-uniform gravitomagnetic field, we should use the following expression, in analogy to the effect of an external magnetic field in non-relativistic quantum mechanics,% as\enlargethispage{0.5cm}
\begin{equation}
\begin{aligned}
\Phi_{\rm m}&=\Phi_{\rm m1}+\Phi_{\rm m2}\\
&=\frac{i\hbar}{m} \mathbf{A}_{\rm g}\cdot \bigtriangledown+\frac{1}{2} \mathbf{A}_{\rm g}\cdot\mathbf{A}_{\rm g}%\\
%&=-\frac{1}{mr}A_{\rm g}\hat{L}_{z}+\frac{1}{2}A_{\rm g}A_{\rm g}
,
% \Phi_{\rm m}&=\Phi_{\rm m1}+\Phi_{\rm m2}\\
% &=-\frac{1}{2m}\mathbf{B}_{\rm g}\cdot\hat{\mathbf{L}}+\frac{1}{8}[(\mathbf{r}\cdot \mathbf{r})(\mathbf{B}_{\rm g}\cdot\mathbf{B}_{\rm g})-(\mathbf{r}\cdot\mathbf{B}_{\rm g})(\mathbf{r}\cdot\mathbf{B}_{\rm g})].
\end{aligned}
\end{equation}%
where $\mathbf{A}_{\rm g}$ is the gravitomagnetic vector potential, related to the gravitomagnetic field as $\mathbf{B}_{\rm g}=\bigtriangledown\times \mathbf{A}_{\rm g}$. For the vector potential component in the azimuthal angular direction $\partial_\phi$, it takes the form
\begin{align} 
\label{eq:Pm00}
    \Phi_{\rm m}&=-\frac{1}{mr}A^\phi_{\rm g}\hat{L}_{z}+\frac{1}{2}A^\phi_{\rm g}A^\phi_{\rm g}.
\end{align}% However, we can approximate the potential with Eq.~\eqref{eq:Pm00} in regions where the variations of gravitomagnetic field are sufficiently small. In this section we use Eq.~\eqref{eq:Pm00} to approximate %as an approximation of 
%the values of the potential in infinitesimal regions where the gravitomagnetic field variations can be neglected.
%
Generally speaking, the linear term $\Phi_{\rm m1}$ is much larger than the quadratic one $\Phi_{\rm m2}$. In this section, we will talk about some possible sources of $\mathbf{A}_{\rm g}$ %$\mathbf{B}_{\rm g}$ 
and their effects on the FDM solitons.%\Mov{ For comparison between different gravitational potentials, we used in this section a different normalisation from Footnote~\ref{ftn:norm}, adjusted to follow Eq.~\eqref{eq:P10}, such that all potentials vanish at radial infinity. As in this section we do not solve the SP system, this change of normalisation does not affect our results.
%}

\subsection{Gravitomagnetic field due to rotation velocity of fuzzy dark matter}
\label{sec:gm0}
The gravitational potential inside an FDM soliton sourced by its own density profile follows the spherically  symmetric potential%is
\begin{equation}
\label{eq:Pe10}
\Phi=-\frac{4\pi G\int_{0}^{r}\rho(R)R^2dR}{r}-4\pi G\int_{r}^{\infty}\frac{\rho(R)R^2dR}{R},
\end{equation}
where we temporarily utilize the normalization $\Phi(r=\infty)=0$ in Sec.~\ref{sec:gm}, as we did not solve the SP system but rather estimate and compare the effect of different potentials%this subsection
.
If this FDM soliton is rotating around the $z-$axis, there should be a non-zero $A^\phi_{\rm g}$ in the $x-y$ plane %gravitomagnetic field $\mathbf{B}_{\rm g}$ \Mov{along $z$ }
according to %the 
gravitoelectromagnetism. To calculate the total $A^\phi_{\rm g}$, in the $x-y$ plane, %$\mathbf{B}_{\rm g}$ 
inside the FDM soliton, we first consider an FDM density shell with radius $R$ rotating around the $z-$axis with a typical linear velocity, in the $x-y$ plane, relative to the FDM soliton's center $v\sim10^{-3}c$. The infinitesimal $dA^\phi_{\rm g}$ % magnitude of the gravitomagnetic \Mov{vector }field 
in the $x-y$ plane %\Mov{, aligned with the $\phi$ direction,} 
produced by this FDM spherical shell verifies%is
\begin{equation} 
dA^\phi_{\rm g}=\begin{cases}
\begin{aligned}
-\frac{4\pi G}{3c^2}\frac{v\rho(R)R^3dR}{r^2},\ \ (r>R),\\
-\frac{4\pi G}{3c^2}v\rho(R)r dR,\ \  (r<R).
\end{aligned}
\end{cases}
% dB_{\rm g}=\begin{cases}
% \begin{aligned}
% \frac{4\pi G}{3c^2}\frac{v\rho(R)R^3dR}{r^3},\ \ (r>R),\\
% -\frac{8\pi G}{3c^2}v\rho(R) dR,\ \  (r<R).
% \end{aligned}
% \end{cases}
\end{equation}%
Then the total $A^\phi_{\rm g}$ %magnitude of the gravitomagnetic field in the $x-y$ plane 
produced by the whole FDM soliton sums up to%is
\begin{equation} 
\begin{aligned}
A^\phi_{g}&=-\frac{4\pi G v}{3c^2}\bigg(\frac{1}{r^2}\int_{0}^{r}\rho(R)R^2\ dR+r\int_{r}^{\infty}\rho(R)\  dR\bigg)\\
&=\frac{4\pi G v}{3c^2}f(r),\label{eq:AphiG}
% B_{g}&=\frac{4\pi G v}{3c^2}\bigg(\frac{1}{r^3}\int_{0}^{r}\rho(R)R^3\ dR-2\int_{r}^{\infty}\rho(R)\  dR\bigg)\\
% &=\frac{4\pi G v}{3c^2}f(r).
\end{aligned}
\end{equation}%
% ???\Mov{We use }%There is 
% an analytic fit to the soliton density from~\cite{Schive:2014dra}
% \begin{equation} 
% \rho(r)\approx\frac{0.019\times\left [m/(10^{-22}{\rm{eV}}/c^2)\right ]^{-2}(r_c/\rm{kpc})^{-4}}{[1+0.091\times(r/r_{c})^2]^8}M_{\odot}\rm{pc}^{-3},\label{eq:densitySP}
% \end{equation}%
% where $r_c$ is the radius of the soliton, \Mov{defined with}
% \begin{equation} 
% M\times r_{c}\approx 2.2\times 10^8 \left [m/(10^{-22} {\rm{eV}}/c^2)\right ]^{-2}M_{\odot}\rm{kpc},
% \end{equation}
% \Mov{and where }%and 
where the density $\rho$ is rescaled from the dimensionless density of the black curve in Fig.~\ref{fig:ge}. The scaling factor $\lambda$ is determined from the comparison between the integration of the black curve and a given soliton mass $M$.
%where the density $\rho$ is rescaled and dimensionalised from the normalised dimensionless density of the black curve in Fig.~\ref{fig:ge} through the soliton density and Eqs.~\eqref{eq:dimLessPsi} and \eqref{eq:PsiTildeScaling}. The scaling factor needed in the rescaling is determined from setting the real soliton mass through Eqs.~\eqref{eq:dimLessM} and \eqref{eq:MtildeScaling}, compared with its observed value. 
T%t
he soliton mass $M$ can be predicted from the halo mass $M_{\rm{halo}}$ according to the soliton-halo mass relation, whether from the version of Ref.~\cite{Schive:2014hza}
\begin{equation} 
\label{eq:r_h}
M \approx 1.25\times 10^{9}\left(\frac{M_{\rm{halo}}}{10^{12}M_{\odot}}\right)^{1/3} \left(\frac{m}{10^{-22}{\rm{eV}}/c^2}\right)^{-1}M_{\odot},
\end{equation}
or following the version of Ref.~\cite{Chan:2021bja}
\begin{equation} 
\label{eq:r_h_n}
\begin{aligned}
M \approx &\beta \left(\frac{m}{8\times10^{-23}\rm{eV}/c^2}\right)^{-3/2}\\
&+\left (\frac{M_{\rm{halo}}}{\gamma}\right )^{\alpha}\left(\frac{m}{8\times10^{-23}\rm{eV}/c^2}\right)^{3(\alpha-1)/2}M_{\odot},
\end{aligned}
\end{equation}%
where $\beta=8.00\times 10^{6}M_{\odot}$, $\gamma=10^{-5.73}M_{\odot}$ and $\alpha=0.515$. The following discussion takes the Milky Way value with $M_{\rm{halo}}=1\times 10^{12} M_{\odot}~\cite{Wang:2019ubx}$ as an example.
%For a\Mov{n} FDM particle rotating in the $x-y$ plane, the magnitude of its angular momentum is $L=mvr$, where its \Mov{typical center of mass }linear velocity %relative to the FDM soliton's center is typically 
%\Mov{is taken as }$v\sim10^{-3}c$.

\begin{figure*}[]
\begin{center}
\subfloat{\includegraphics[width=1\textwidth]{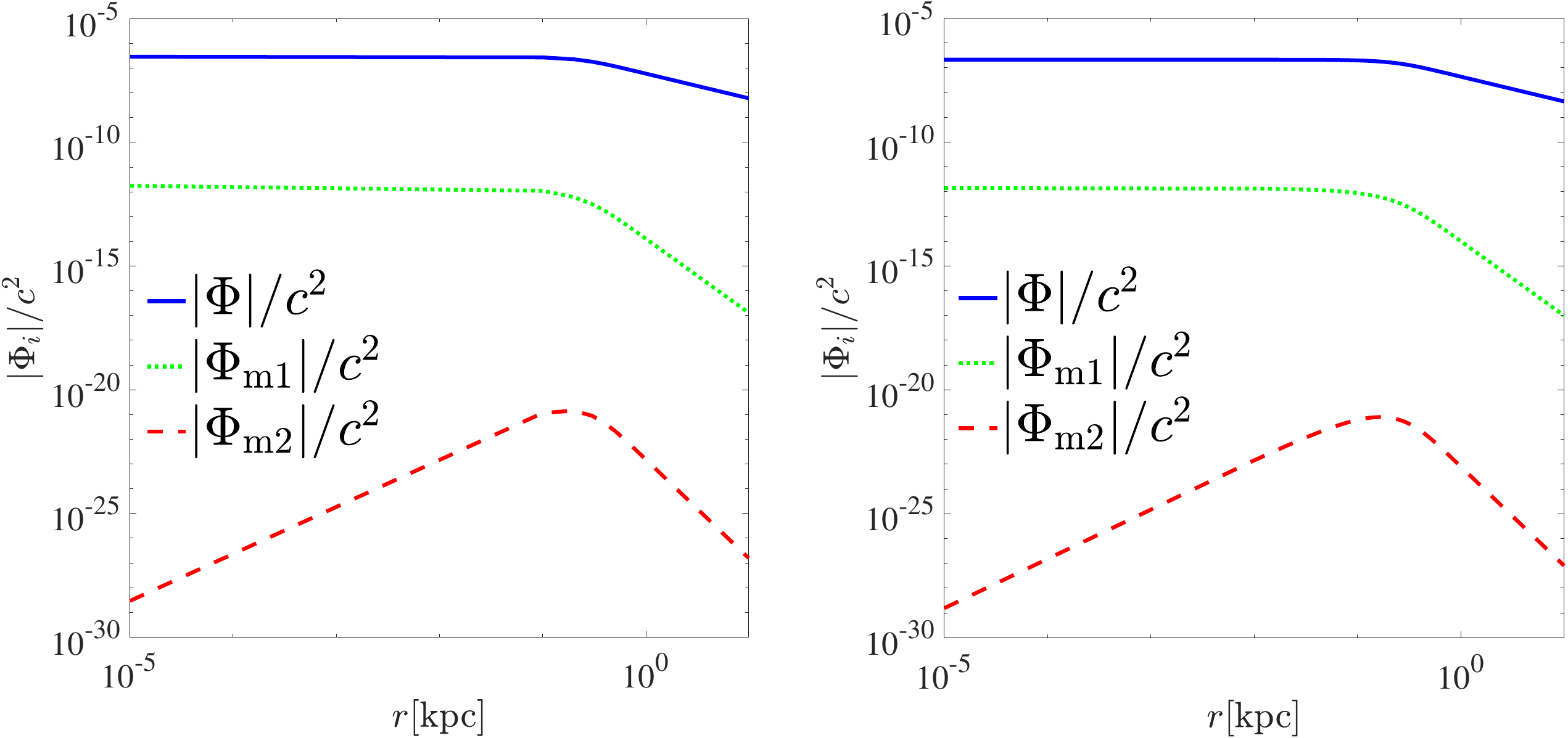}}
\end{center}
\captionsetup{justification=raggedright}
\caption{Gravitational %The gravitational 
potentials %of 
$|\Phi|$ (Eq.~\ref{eq:Pe10}), $|\Phi_{\rm m1}|$ (Eq.~\ref{eq:Pm11}) and $|\Phi_{\rm m2}|$ (Eq.~\ref{eq:Pm12}) in the $x-y$ plane, where the halo mass is similar to the Milky Way's, $M_{\rm{halo}}\approx1\times10^{12}M_{\odot}$~\cite{Wang:2019ubx}, and the FDM particle has the typical mass $m=10^{-22}\rm{eV}/c^2$. In the left panel, we used the soliton-halo mass relation from Eq.~\eqref{eq:r_h}, while the right panel used Eq.~\eqref{eq:r_h_n}.}
\label{fig:P1}
\end{figure*}

The angular momentum of each particle of the soliton being identical, and in the $z$ direction, we compute its magnitude by dividing the FDM shells angular momentum, integrated over the soliton, %soliton angular momentum 
by the number of FDM particles in the soliton $N=M/m$, according to
\begin{equation} 
%\label{eq:L}
L_z=\frac{1}{N}\int_{0}^{\infty}\frac{8\pi}{3}\rho(R)R^3vdR.\label{eq:AngMom}
\end{equation}%\Mov{\bf [I compute this should be $L=2mv\left<r\right>$]}
According to Eq.~(\ref{eq:Pm00}), the linear gravitational potential in the $x-y$ plane yields%is 
\begin{equation} 
\label{eq:Pm11}
\Phi_{\rm m1}=-\frac{1}{mr}A^\phi_{\rm g}L_z,
% \Phi_{\rm m1}=-\frac{1}{2m}B_{\rm g}L=-\frac{2\pi\times10^{-6}}{3}Gf(r)r,
\end{equation}%\Mov{\bf [actually with the above L I compute $\Phi_{\rm m1}=-\frac{4\pi\times10^{-6}}{3}Gf(r)\left<r\right>$!]}
and the quadratic gravitational potential in the $x-y$ plane can be written as%is
\begin{equation} 
\label{eq:Pm12}
\Phi_{\rm m2}=\frac{1}{2}A^\phi_{\rm g}{}^{\,2}=\frac{8\pi^2\times10^{-6}}{9}\frac{G^2}{c^2}f(r)^2.
% \Phi_{\rm m2}=\frac{1}{8}B_{\rm g}^2r^2=\frac{2\pi^2\times10^{-6}}{9}\frac{G^2}{c^2}f(r)^2r^2.
\end{equation}%
The gravitational potentials $\Phi$ (Eq.~\ref{eq:Pe10}), $\Phi_{\rm m1}$ (Eq.~\ref{eq:Pm11}) and $\Phi_{\rm m2}$ (Eq.~\ref{eq:Pm12}) include some nontrivial %complicated 
integrals, which obscures their direct comparison% so it's not easy to compare them directly
. 
For clarity, we plot the gravitational potentials %of 
$|\Phi|$, $|\Phi_{\rm m1}|$ and $|\Phi_{\rm m2}|$ %in the $x-y$ plane 
in Fig.~\ref{fig:P1}. We find $\Phi$ to dominate %is dominant 
over $\Phi_{\rm m1}$ and $\Phi_{\rm m2}$ in the $x-y$ plane and thus that the latter two can be neglected. 

The above introduction of angular momentum was based on a background soliton obeying Eq.~\eqref{eq:sp00}, without rotation, that yields the density of the black curve in Fig.~\ref{fig:ge}%Eq.~\eqref{eq:densitySP}
. The effect of rotation is modeled in Eq.~\eqref{eq:Pm00} and has been shown to be negligible compared with the soliton's self-gravitating potential. 
%We conclude that introduction of angular momentum, with FDM shells linear velocity at least up to order $10^{-3} c$, should not violate the soliton stability, and thus the soliton is allowed to rotate under its own gravitomagnetic field.
%
% In general the soliton is not expected rotate. The above calculation shows that even if it were, the effect of rotation are negligible and therefore should not be taken into account. Everything behaves as if the soliton does not rotate.

% \begin{figure*}[]
% \begin{center}
% \subfloat{\includegraphics[width=0.8\textwidth]{rho.jpg}}
% \end{center}
% \captionsetup{justification=raggedright}
% \caption{}
% \label{fig:ge}
% \end{figure*}  

\subsection{Gravitomagnetic field due to a supermassive black hole spin}
\label{sec:gm1}
In this section, we consider the effect %of 
not only of the gravitoelectric field but also of the angular momentum %gravitomagnetic field 
from% %the angular momentum of the central supermassive BH
 the central supermassive BH %angular momentum 
on a soliton rotating in a similar way as Sec.~\ref{sec:gm0}. In this case we express potentials in terms of a dimensionless distance in units of %Firstly we relate 
the Schwarzschild radius $R_{\rm S}=\frac{2GM_{\rm bh}}{c^2}$. To do so we define it for % to 
an arbitrary radial distance $r$ as $n=r/R_{\rm S}$. 
%For example, %for the supermassive BH at the center of the Milky Way\Mov{ has a mass of }%, its mass is $M_{\rm bh}\approx4\times10^6M_{\odot}$ and Schwarzschild radius %is $R_{\rm S}\approx1.2\times10^{10} {\rm m}\approx4\times 10^{-7}{\rm pc}$.
The %n the 
gravitational potential sourced by a central supermassive BH follows%is
\begin{equation} 
\label{eq:Pe20}
\Phi_{\rm e}=-\frac{GM_{\rm bh}}{r}=-\frac{c^2}{2n}.
\end{equation}
If this supermassive BH is spinning, the magnitude of its angular momentum is $L_{\rm bh}=\chi\frac{GM_{\rm bh}^2}{c}$, where $\chi$ is a dimensionless spin parameter. According to %the 
gravitoelectromagnetism, the gravitomagnetic vector potential field $\mathbf{A}_{\rm g}$ depends %is dependent 
on the angular momentum $\mathbf{L}_{\rm bh}$, which points along the $z$ direction, as
\begin{equation}
\label{eq:B} 
\mathbf{A}_{\rm g}=-\frac{G}{2c^2r^3}\mathbf{L}_{\rm bh}\times\mathbf{r}. 
% \mathbf{B}_{\rm g}=\frac{G}{2c^2}\frac{\mathbf{L}_{\rm bh}-3(\mathbf{L}_{\rm bh}\cdot\mathbf{r}/r)\mathbf{r}/r}{r^3}.
\end{equation}%
The corresponding %refore, the 
magnitude of the gravitomagnetic vector potential field $A^\phi_{\rm g}$ due to this spinning massive BH in the $x-y$ plane reads%is 
\begin{equation} 
A^\phi_{\rm g}=-\frac{G}{2c^2}\frac{L_{\rm bh}}{r^2}=-\frac{\chi}{8n^2}c.% B_{\rm g}=\frac{G}{2c^2}\frac{L_{\rm bh}}{r^3}=\frac{\chi}{8n^3}\frac{c}{R_{\rm S}}.
\end{equation}

\begin{figure*}[]
\begin{center}
\subfloat{\includegraphics[width=1\textwidth]{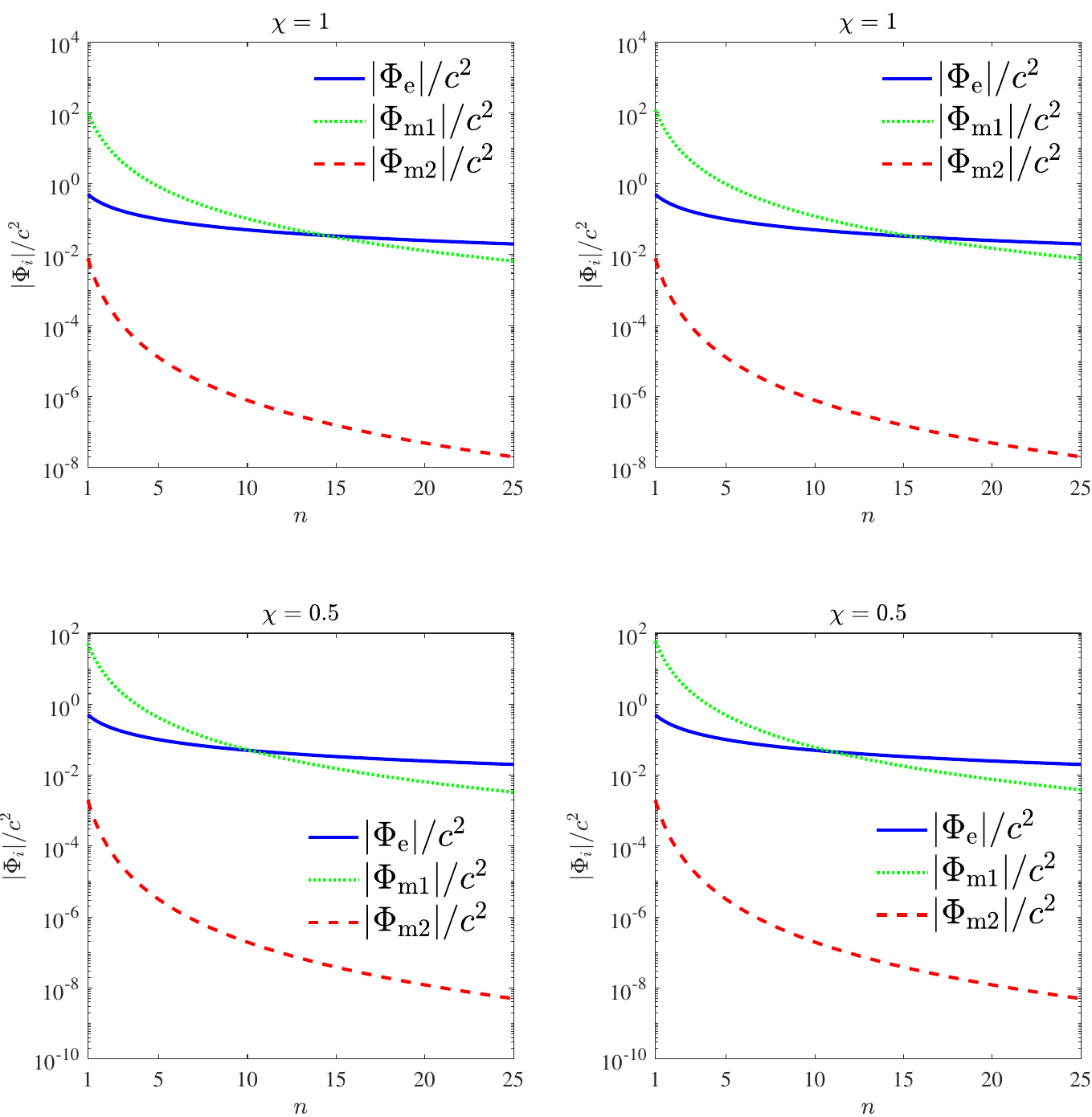}}
\end{center}
\captionsetup{justification=raggedright}
\caption{Gravitational %The gravitational 
potentials %of 
$|\Phi_{\rm e}|$, $|\Phi_{\rm m1}|$ and $|\Phi_{\rm m2}|$ in the $x-y$ plane as a function of $n$ from Eqs%calculated by Eq
.~(\ref{eq:Pe20}, \ref{eq:Pm21} and \ref{eq:Pm22}) respectively, where $n\geq1$ marks that %means 
the radial distance is larger than the Schwarzschild radius of the supermassive BH. In the top %The left 
subplots we set %is plotted with 
$\chi=1$ while in the bottom %and the right 
subplots, % is plotted with 
$\chi=0.5$. Concomitantly, the left panels used the soliton-halo mass relation from Eq.~\eqref{eq:r_h}, while the right panels used Eq.~\eqref{eq:r_h_n}.} 
\label{fig:P2}
\end{figure*}

For an FDM particle, we can obtain its $z$ direction %rotating in the $x-y$ plane, the magnitude of its 
angular momentum %is 
by Eq.~\eqref{eq:AngMom}%$L=mvr$
, where %its linear velocity relative to the central supermassive BH is typically 
$v\sim10^{-3}c$ and $\rho$ is rescaled from the green curve in Fig.~\ref{fig:ge}. $\lambda$ is determined from the comparison between the integration of the green curve and $M$ (Eq.~\ref{eq:r_h} or \ref{eq:r_h_n}). Meanwhile, $M_{\rm bh}$ is also rescaled from $\tilde{M}_{\rm bh}=0.5$ with the same $\lambda$.
According to Eq.~(\ref{eq:Pm00}), the linear gravitational potential in the $x-y$ plane follows%is 

\begin{equation} 
\label{eq:Pm21}
\Phi_{\rm m1}=-\frac{1}{mr}A^\phi_{\rm g}L_z=\frac{1}{m}\frac{\chi}{8n^3}\frac{c}{R_{\rm S}}L_z, 
% \Phi_{\rm m1}=-\frac{1}{2m}B_{\rm g}L%\sim-\frac{\chi}{16n^2}\times 10^{-3}c^2, 
\end{equation}%
while the corresponding %and the 
quadratic gravitational potential writes%in the $x-y$ plane is
\begin{equation}
\label{eq:Pm22}
\Phi_{\rm m2}=\frac{1}{2}A^\phi_{\rm g}{}^{\,2}=\frac{\chi^{2}}{128n^4}c^2.
% \Phi_{\rm m2}=\frac{1}{8}B_{\rm g}^2r^2=\frac{\chi^{2}}{512n^4}c^2.
\end{equation}%
Comparing the gravitational potentials from Eqs%calculated by Eq
.~(\ref{eq:Pe20}, %Eq.~(
\ref{eq:Pm21} and %Eq.~(
\ref{eq:Pm22}), we find $\Phi_{\rm e}$ %is dominant 
and %over 
$\Phi_{\rm m1}$ to dominate over %and 
$\Phi_{\rm m2}$ in the $x-y$ plane and thus that the latter %two 
can be neglected for $n\geq1$ (ouside the SMBH horizon). For clarity, we plot the gravitational potentials %of 
$|\Phi_{\rm e}|$, $|\Phi_{\rm m1}|$ and $|\Phi_{\rm m2}|$ in the $x-y$ plane as a function of $n$ in Fig.~\ref{fig:P2}, where $n\geq1$ marks that %means 
the radial distance is larger than the Schwarzschild radius of the supermassive BH.
We found that $\Phi_{\rm m1}$ is comparable to $\Phi_{\rm e}$ for radii within a decade of the Schwarzschild radius. According to Sec.~\ref{sec:ge}, this will further modify the soliton profile obtained from Eq.~\eqref{eq:sp10}. Then, following Eq.~\eqref{eq:AngMom}, the angular momentum of the FDM particle should be affected, which in turn should modify $\Phi_{\rm m1}$ through Eq.~\eqref{eq:Pm21}%\Ke{(31)}
. This loop reveals that our method cannot yield safely an evaluation of the potential for the gravitomagnetic field induced by the spin of a supermassive BH. To do so, the spherical symmetric approach of $\Phi$ (Eq.~\ref{eq:Pe10}) is insufficient to account for the cylindrical symmetric gravitomagnetic potential $\Phi_{\rm m1}$. Solving the cylindrical symmetric system (Eq.~\ref{eq:sp20}) lies beyond the scope of the present paper.
% For %these 
% FDM particles not located in the $x-y$ plane\Mov{, although their angular momentum remains identical, the effect of $\Phi_{\rm m1}$ is much smaller there, due to the smaller magnitude of $B_{\rm g}$, and} %or located in the $x-y$ plane but 
% %not rotating around the central supermassive BH, 
% $\Phi_{\rm e}$ %is 
% further 
% domina\Mov{tes all potentials}%nt
% .\Mov{ The soliton is thus allowed to stably rotate in the presence of a central supermassive BH, for FDM shells %angular momenta 
% with linear velocity at least up to order $10^{-3}c$.}

\subsection{Gravitomagnetic field due to the orbital motion of a supermassive
black hole binary}
\label{sec:gm2}
In this section, we consider the effect %of 
not only of the gravitoelectric field but also of the angular momentum from a supermassive BH binary %angular momentum 
on a %not rotating 
soliton rotating in a similar way as Sec.~\ref{sec:gm0}%with an angular momentum fluctuation introduced by one FDM particle
. For simplicity, we consider an equal mass supermassive BH binary in a circular orbit in the $x-y$ plane. These two supermassive BHs share the same mass $M_{\rm bh}$ and have a separation distance $S=sR_{\rm S}$ between each other, where $s$ is a dimensionless parameter. The %n the 
magnitude of the angular momentum for such %of this 
binary is
\begin{equation} 
L_{\rm bbh}=I_{\rm bbh}\Omega_{\rm bbh}=\sqrt{s}\frac{GM_{\rm bh}^2}{c},
\end{equation}
where the moment of inertia of this binary reduces to  %is 
$I_{\rm bbh}=\frac{1}{2} M_{\rm bh} S^{2}$ and its %the 
angular frequency follows %of this binary is 
$\Omega_{\rm bbh}=\frac{1}{\sqrt{s}}\frac{c}{S}$.
According to Eq.~(\ref{eq:B}), the magnitude of the gravitomagnetic field due to this binary in the $x-y$ plane yields%is 
\begin{equation} 
A^\phi_{\rm g}=-\frac{G}{2c^2}\frac{L_{\rm bbh}}{r^2}=-\frac{\sqrt{s}}{8n^2}c.
% B_{\rm g}=\frac{G}{2c^2}\frac{L_{\rm bbh}}{r^3}=\frac{\sqrt{s}}{8n^3}\frac{c}{R_{\rm S}}
\end{equation}

\begin{figure*}[]
\begin{center}
\subfloat{\includegraphics[width=1\textwidth]{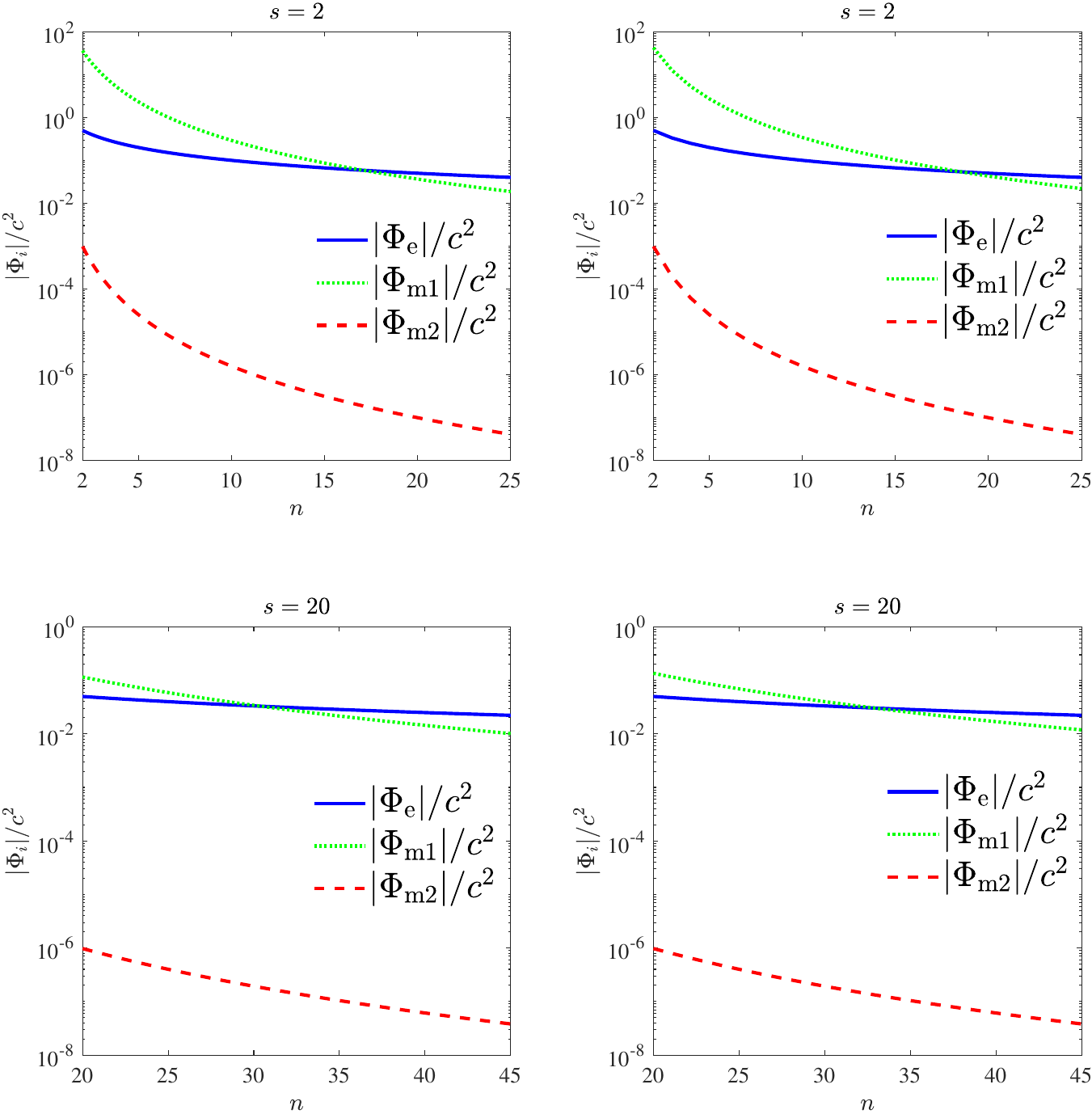}}
\end{center}
\captionsetup{justification=raggedright}
\caption{Gravitational %The gravitational 
potentials %of 
$|\Phi_{\rm e}|=\frac{c^2}{n}$, $|\Phi_{\rm m1}|$ (Eq.~\ref{eq:Pm31}) and $|\Phi_{\rm m2}|$ (Eq.~\ref{eq:Pm32}) in the $x-y$ plane as a function of $n$, where $n\geq1+\frac{s}{2}$ marks that %means 
the radial distance is larger than the Schwarzschild radius of each supermassive BH plus half the binary separation, $R_{\rm S}+\frac{S}{2}$. In the top %The left 
subplots we set %is plotted with 
$s=2$ while in the bottom %and the right 
subplots, % is plotted with 
$s=20$. Concomitantly, the left panels used the soliton-halo mass relation from Eq.~\eqref{eq:r_h}, while the right panels used Eq.~\eqref{eq:r_h_n}.}
\label{fig:P3}
\end{figure*}

In addition, for an %Also for a 
FDM particle, we can obtain its $z$ direction % rotating in the $x-y$ plane, the magnitude of its 
angular momentum by Eq.~\eqref{eq:AngMom}%is $L=mvr$
, where %its linear velocity relative to this binary is typically 
$v\sim10^{-3}c$ and $\rho$ is rescaled from the green curve in Fig.~\ref{fig:ge}. $\lambda$ is determined from the comparison between the integration of the green curve and $M$ (Eq.~\ref{eq:r_h} or \ref{eq:r_h_n}). Meanwhile, $2M_{\rm bh}$ is also rescaled from $\tilde{M}_{\rm bh}=0.5$ with the same $\lambda$.
According to Eq.~(\ref{eq:Pm00}), the linear gravitational potential in the $x-y$ plane follows%is 
\begin{equation} 
\label{eq:Pm31}
\Phi_{\rm m1}=-\frac{1}{mr}A^\phi_{\rm g}L_z=\frac{1}{m}\frac{\sqrt{s}}{8n^3}\frac{c}{R_{\rm S}}L_z, 
% \Phi_{\rm m1}=-\frac{1}{2m}B_{\rm g}L%\sim-\frac{\sqrt{s}}{16n^2}\times10^{-3}c^2,
\end{equation}%
while the corresponding %and the 
quadratic gravitational potential writes%in the $x-y$ plane is
\begin{equation} 
\label{eq:Pm32}
\Phi_{m2}=\frac{1}{2}A^\phi_{\rm g}{}^{\,2}=\frac{s}{ 128n^4}c^2.
% \Phi_{m2}=\frac{1}{8}B_{\rm g}^2 r^2=\frac{s}{ 512n^4}c^2.
\end{equation}%
We plot the gravitational potentials %of 
$|\Phi_{\rm e}|=\frac{c^2}{n}$, $|\Phi_{\rm m1}|$ (Eq.~\ref{eq:Pm31}) and $|\Phi_{\rm m2}|$ (Eq.~\ref{eq:Pm32}) in the $x-y$ plane as a function of $n$ in Fig.~\ref{fig:P3}, where $n\geq1+\frac{s}{2}$ marks that %means 
the radial distance is larger than the Schwarzschild radius of each supermassive BH plus half the binary separation, $R_{\rm S}+\frac{S}{2}$. %
We found that $\Phi_{\rm m1}$ is comparable to $\Phi_{\rm e}$ for radii within a decade of the Schwarzschild radius. Similarly to subsec.~\ref{sec:gm1}, this will further modify the soliton profile obtained from Eq.~\eqref{eq:sp10}. Then, following Eq.~\eqref{eq:AngMom}, the angular momentum of the FDM particle should be affected, which in turn should modify $\Phi_{\rm m1}$ through Eq.~\eqref{eq:Pm31}%\Ke{(35)}
. Therefore, the spherical symmetric approach of $\Phi$ (Eq.~\ref{eq:Pe10}) is also insufficient to account for the cylindrical symmetric gravitomagnetic potential $\Phi_{\rm m1}$.

% We find $\Phi_{\rm e}$ \Mov{to dominate }%is dominant 
% over $\Phi_{\rm m1}$ and $\Phi_{\rm m2}$ in the $x-y$ plane and \Mov{thus that }the latter two can be neglected for $n\geq1+\frac{s}{2}$\Mov{. It even more dominates over the potentials felt by }%, let alone these 
% FDM particles %\Mov{either }
% not located in the $x-y$ plane\Mov{ since, although their angular momentum remains identical, the effect of $\Phi_{\rm m1}$ there is much smaller, due to the smaller magnitude of $B_{\rm g}$. The soliton is thus allowed to stably rotate in the presence of a central supermassive BH binary, for %FDM particle angular momentum with 
% FDM shells linear velocity at least up to order $10^{-3}c$.} %or located in the $x-y$ plane but 
%not rotating around this binary.
%
% ----------------------
%
% For FDM particles not located in the $x-y$ plane\Mov{, although their angular momentum remains identical, the effect of $\Phi_{\rm m1}$ is much smaller there, due to the smaller magnitude of $B_{\rm g}$, and} %or located in the $x-y$ plane but 
% %not rotating around the central supermassive BH, 
% $\Phi_{\rm e}$ %is 
% further 
% domina\Mov{tes all potentials}%nt
% .\Mov{ The soliton is thus allowed to stably rotate in the presence of a central supermassive BH, for angular momenta with linear velocity at least up to order $10^{-3}c$.}

\section{Effects of the extreme density-ratio %of 
soliton binary}
\label{sec:edr}
\begin{figure*}[]
\begin{center}
\subfloat{\includegraphics[width=1\textwidth]{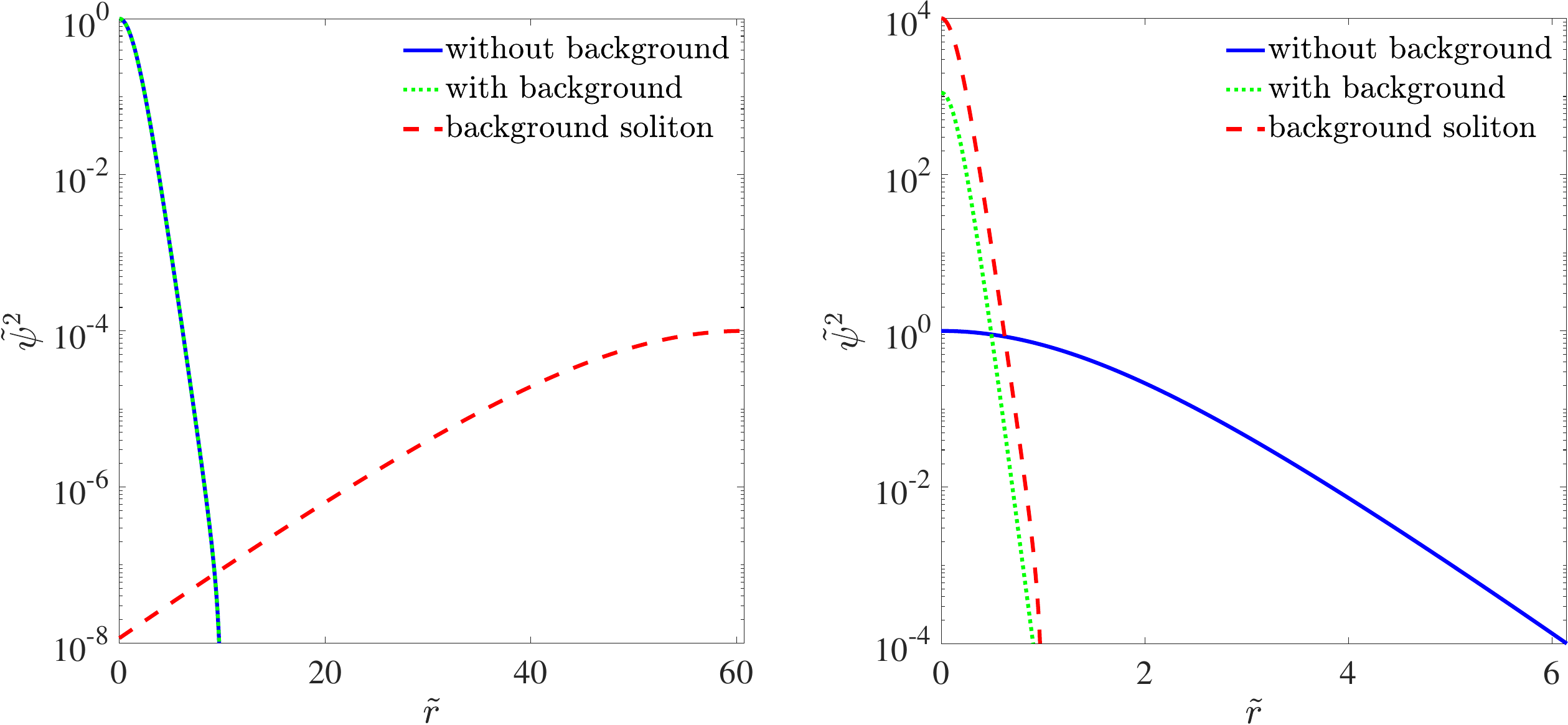}}

\end{center}
\captionsetup{justification=raggedright}
\caption{Density %The density 
profile $\tilde{\psi}^2(\tilde{r})$ of each soliton for stage 1 (left) and stage  2 (right), for the case of Sec.~\ref{sec:edr}.}
\label{fig:2s}
\end{figure*}

The case of a system of two solitons, with %For the cases where the 
density-ratio %of two solitons is 
$\sim1$, %this system 
obviously does not feature %the 
spherical symmetry and could require %can be studied by 
numerical simulations~\cite{Paredes:2015wga,Edwards:2018ccc,Munive-Villa:2022nsr}.
On the contrary, %for 
the extreme case where the two solitons density-ratio %of two solitons is 
$\gtrsim10^4$ presents %, there are 
two stages when this system do feature an approximate and local spherical symmetry: in stage 1) the higher density, smaller soliton lies immersed within %with higher density is just immerse in 
the boundary of the lower density, thus larger and relatively flatter, soliton %with lower density 
which %is relatively flat and 
can be considered as a background; in stage 2) the lower density, larger soliton %with lower density 
shares the same center as %with 
the higher density, smaller and more stable, soliton %with higher density 
which %are more stable and 
can then be considered as a background. The collisional dynamics include these two stages and we can still use the shooting method to deal with them. Finally, we can use
the exact SP system (Eq.~\ref{eq:sp00}) to describe the background soliton and the following variant of the exact SP system to describe the soliton with background in either of %during 
the above two stages
\begin{equation}
\label{eq:sp30}
\begin{cases}
\begin{aligned}
&i\hbar\frac{\partial \Psi}{\partial t}=\left(-\frac{\hbar^2}{2m}\nabla^2+m\Phi\right)\Psi ,\\
&\nabla^2 \Phi=4\pi G (|\Psi|^2+ \rho_{\rm bg}),\\
\end{aligned}
\end{cases}
\end{equation}
whose dimensionless version is
\begin{equation}
\label{eq:sp31}
\begin{cases}
\begin{aligned}
\frac{\partial^2 (\tilde{r}\tilde{\psi})}{\partial \tilde{r}^2}=2\tilde{r}(\tilde{\Phi}-\tilde{\gamma})\tilde{\psi},\\
\frac{\partial^2 (\tilde{r}\tilde{\Phi})}{\partial \tilde{r}^2}=\tilde{r}(\tilde{\psi}^2+\tilde{\rho}_{\rm bg}),
\end{aligned}
\end{cases}
\end{equation}
where $\rho_{\rm bg}$ is the density of the background soliton and its dimensionless counterpart is
\begin{equation}
\tilde{\rho}_{\rm bg}\equiv\left(\frac{\hbar\sqrt{4 \pi G}}{mc^2}\right)^2\rho_{\rm bg}.
\end{equation}
Since this variant SP system (Eq.~\ref{eq:sp31}) also obeys %has 
the scaling symmetry when
\begin{align}
%&\tilde{\psi}  \longrightarrow  \lambda \tilde{\psi}, \\
%&\tilde{r}  \longrightarrow  \lambda^{-1/2} \tilde{r},\\
%&\tilde{\Phi}  \longrightarrow  \lambda \tilde{\Phi}, \\
%&\tilde{\gamma}  \longrightarrow  \lambda \tilde{\gamma}, \\
%&\tilde{M} \longrightarrow  \lambda^{1/2} \tilde{M},\\
&\tilde{\rho}_{\rm bg} \longrightarrow  \lambda^{2} \tilde{\rho}_{\rm bg},
\end{align}
we can choose to set up the background soliton with $\lambda=10^{-2}$ for stage 1 and with $\lambda=10^{2}$ for stage 2. Given the background density $\rho_{\rm bg}$, the soliton with background can be found by solving Eq.~(\ref{eq:sp31}) with the shooting method. By assuming that its density profile $\tilde{\psi}^2(\tilde{r})$ can be different, while %is changing but 
its dimensionless mass $\tilde{M}$ remains %is 
fixed as the value listed in the first row of Tab.~\ref{tb:ge}, we get the eigenvalue $\tilde{\gamma}=0.6495$ for stage 1 and the eigenvalue $\tilde{\gamma}=69.2491$ for stage 2. The %For stage 1, the 
density profile $\tilde{\psi}^2(\tilde{r})$ for %of 
each soliton is shown, for stage 1, in the left subplot, while it is shown, for stage 2, % of Fig.~\ref{fig:2s}; for stage 2, each soliton's $\tilde{\psi}^2(\tilde{r})$ is shown 
in the right subplot of Fig.~\ref{fig:2s}.
We %can 
find that the higher density, smaller soliton remains %with higher density is 
almost un%not 
affected by the background soliton for stage 1 while %and 
the lower density, larger soliton %with lower density 
shrinks dramatically for stage 2.
%\Mov{Following this dramatic evolution after }%
After stage 2, %there are 
two comparable solitons remain, corresponding to a smaller density-ratio case where numerical simulations could be required% and the system becomes complicated whose further evolution should be studied by numerical simulation
.

\section{Effects of the ellipsoidal baryon profile}
\label{sec:bar}
Instead of a background soliton shown in section~\ref{sec:edr}, we now consider a background baryon profile% in this section
. More precisely, we consider an ellipsoidal baryon profile fitted on %similar to 
the Milky Way's~\cite{Iocco:2015xga,Lin:2019yux} three baryonic density components
\begin{equation}
\rho_{\rm bg}(x,y,z)=\rho_{\rm bulge}(x,y,z)+\rho_{\rm disk}(x,y,z)+\rho_{\rm gas}(x,y,z),
\end{equation}
where $\rho_{\rm bulge}$ is a triaxial and bar-shaped bulge of stars at the inner few kpc of the Milky Way~\cite{Lopez-Corredoira:2005anz,Ryu:2008dx}, $\rho_{\rm disk}$ represents the galactic disk density profile and $\rho_{\rm gas}$ gives the diffuse gas component of the Milky Way which takes the form of molecular, atomic and ionised hydrogen and heavier elements. The bulge can be 
fitted by
\begin{equation} 
\begin{aligned}
&\rho_{\rm bulge}(x,y,z)=9.9M_{\odot}{\rm{pc}}^{-3}\times\\
&\exp\left\{-\left[x^2+\left(\frac{y}{0.49}\right)^2+\left(\frac{z}{0.37}\right)^2\right]^{1/2}/740{\rm{pc}}\right\}.
\end{aligned}
\end{equation}
The %the 
disk density, because of its observed value $\rho_{\rm disk}\approx0.05M_{\odot}{\rm{pc}}^{-3}$~\cite{Lopez-Corredoira:2005anz}, can be neglected compared with the bulge density. On the contrary, the diffuse gas density % but 
$\rho_{\rm gas}$,  is 
a non-negligible part of the baryons in the Milky Way% in 
%the form of molecular, atomic and ionised hydrogen and heavier elements
. Here we assume that $\rho_{\rm gas}\approx m_{H_2}\left \langle n_{H_{2}} \right \rangle$, where $m_{H_2}$ is the mass of molecular hydrogen and $\left \langle n_{H_{2}} \right \rangle$ is its %the 
number density %of molecular hydrogen 
in central molecular zone~\cite{Ferriere:2007yq}, fitted by
\begin{equation} 
\begin{aligned}
&\left \langle n_{H_{2}} \right \rangle=150 {\rm{cm}}^{-3}\times\\
&\exp\left\{-\left[\frac{\sqrt{x^2+(2.5y)^2}-125{\rm{pc}}}{137{\rm{pc}}}\right]^4 
\times \exp\left[-\left(\frac{z}{18{\rm{pc}}}\right)^2\right]\right\}.
\end{aligned}
\end{equation}
For this non-spherical system, the shooting method fails %is failed 
and numerical simulations should be performed.
However, we propose %As 
an approximation in which %, however, 
we consider three spherically averaged background baryon profiles respectively: $\rho_{\rm bg}^x(r)=\rho_{\rm bg}(x=r,y=0,z=0)$, $\rho_{\rm bg}^y(r)=\rho_{\rm bg}(x=0,y=r,z=0)$ and $\rho_{\rm bg}^z(r)=\rho_{\rm bg}(x=0,y=0,z=r)$. The %n the 
shooting method can therefore be applied to these spherical approximations %still works for 
$\rho_{\rm bg}^i(r)$. Should %If 
the results by each %different 
$\rho_{\rm bg}^i(r)$ be too markedly %are very 
different, it would indicate that %means 
the original $\rho_{\rm bg}(x,y,z)$ is exceedingly %too 
non-spherical and unable to be described by the approximation $\rho_{\rm bg}^i(r)$. 

The way we %To 
find the equilibrium configurations, following the  procedure of %as shown in 
section~\ref{sec:ge}, %we usually 
uses an % the 
arbitrary normalization $\tilde{\psi}(\tilde{r}=0)=1$ for which %. It means that 
the normalized soliton mass $\tilde{M}=\int_0^\infty \tilde{r}^2\tilde{\psi}^2d\tilde{r}$ %is 
deviates from the true %\Ke{real?} 
dimensionless mass by a scaling factor $\lambda$. For the Milk Way, the soliton physical mass $M\approx1.25\times10^{9}M_{\odot}$ or $M\approx1.16\times10^{9}M_{\odot}$ is obtained with the soliton-halo mass relation Eq.~\eqref{eq:r_h} or \eqref{eq:r_h_n}.
On the other hand, taking this scaling factor into consideration, $\tilde{\rho}_{\rm bg}^i \rightarrow\lambda^{2}\tilde{\rho}_{\rm bg}^i$ modifies %means 
the background baryon profile% is modified
. To set %fix 
both %of 
the soliton mass and the background baryon profile as the Milk Way's, we introduce an {\it a priori} scaling factor %in advance 
and obtain a new dimensionless variant as
\begin{equation}
\label{eq:sp41}
\begin{cases}
\begin{aligned}
&\frac{\partial^2 (\tilde{r}\tilde{\psi})}{\partial \tilde{r}^2}=2\tilde{r}(\tilde{\Phi}-\tilde{\gamma})\tilde{\psi},\\
&\frac{\partial^2 (\tilde{r}\tilde{\Phi})}{\partial \tilde{r}^2}=\tilde{r}(\tilde{\psi}^2+\lambda^{-2}_0\tilde{\rho}_{\rm bg}^i),
\end{aligned}
\end{cases}
\end{equation}
where $\lambda^{-2}_0$ is a parameter unaffected by %and not changing with 
the scaling factor $\lambda$, such that %.
%Therefore, 
this system retains %still has 
the scaling symmetry with $\lambda^{-2}_0\tilde{\rho}_{\rm bg}^i \rightarrow\lambda^{2}\lambda^{-2}_0\tilde{\rho}_{\rm bg}^i$.
Given an initial $\lambda_0$, we can obtain the normalized soliton profile $\tilde{\psi}^2$. The %And then we can obtain the 
scaling factor $\lambda$ can then be computed by requiring the soliton mass to be the Milk Way's. If $\lambda/\lambda_{0}\neq1$, we replace the initial $\lambda_0$ with the computed %newest 
$\lambda$ and repeat the former two
steps. We stop the iteration when 
$\lambda/\lambda_{0}-1<0.0001$%, we stop the iteration
.

\begin{figure*}[]
\begin{center}
\subfloat{\includegraphics[width=0.7\textwidth]{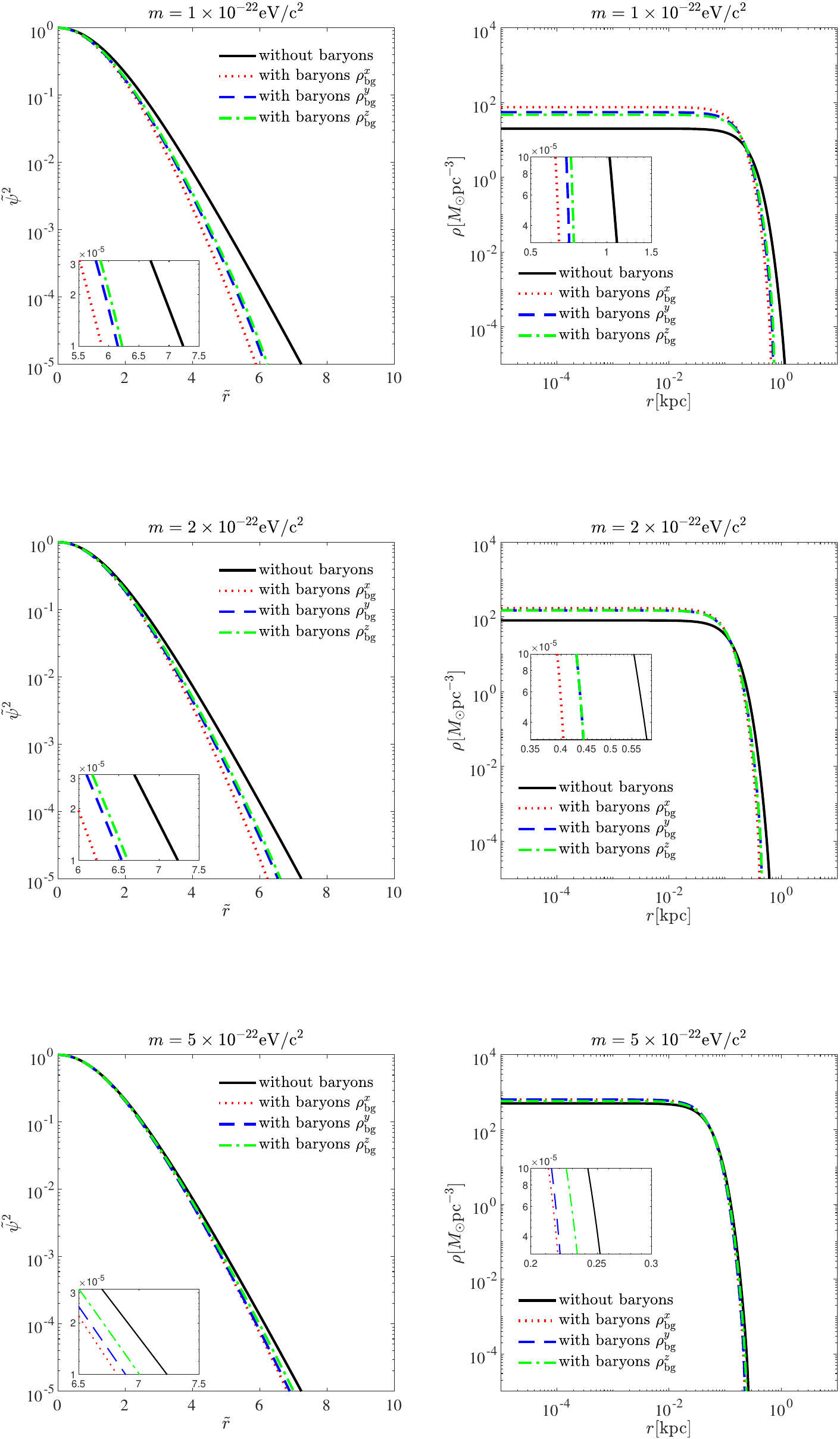}}
\end{center}
\captionsetup{justification=raggedright}
\caption{Density %The density 
profiles of the ground state solutions with different background baryon profile $\tilde{\rho}_{\rm bg}^i$ for various FDM masses %and the mass of FDM 
$m$ and using the soliton-halo mass relation Eq.~\eqref{eq:r_h}: %the 
first row, % is plotted with 
$m=1\times10^{-22}\rm{eV}/c^2$, %the 
second row, % is plotted with 
$m=2\times10^{-22}\rm{eV}/c^2$ and %the 
third row, % is plotted with 
$m=5\times10^{-22}\rm{eV}/c^2$; %the 
first column gives %is the 
normalized density profile while %and 
the second column represents %is 
the physical density profile; %the 
black curves mark pure FDM solitons %is plotted 
without the background baryon profile, %the 
red dotted curves show the effect of %is plotted with 
$\tilde{\rho}_{\rm bg}^x$, %the 
blue dashed curves, that of % is plotted with 
$\tilde{\rho}_{\rm bg}^y$ and %the 
green dash-dotted curves, of % is plotted with 
$\tilde{\rho}_{\rm bg}^z$.}
\label{fig:prof1}
\end{figure*}

\begin{figure*}[]
\begin{center}
\subfloat{\includegraphics[width=0.7\textwidth]{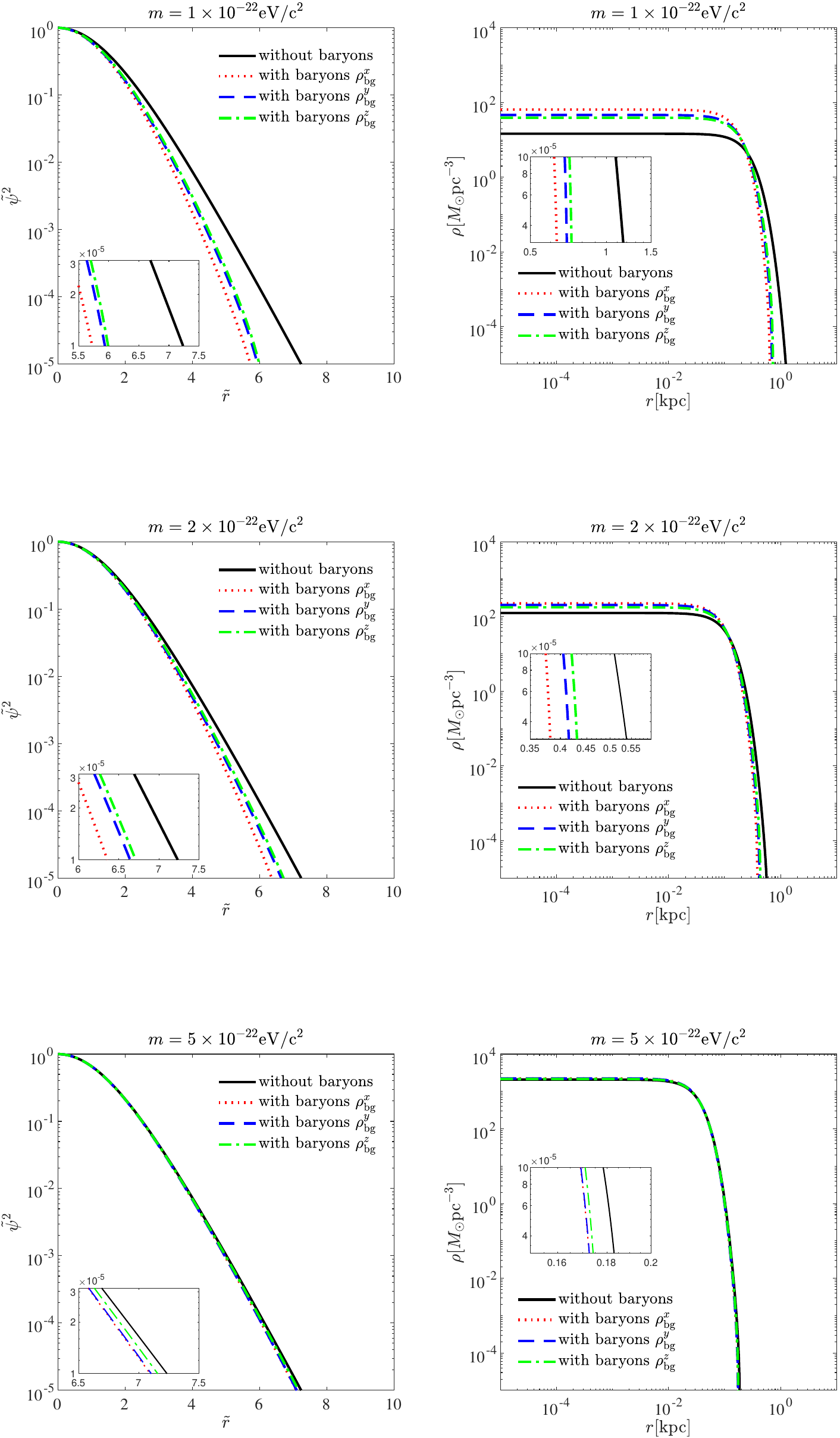}}
\end{center}
\captionsetup{justification=raggedright}
\caption{Density %The density 
profiles of the ground state solutions with different background baryon profile $\tilde{\rho}_{\rm bg}^i$ for various FDM masses %and the mass of FDM 
$m$ and using the soliton-halo mass relation Eq.~\eqref{eq:r_h_n}: %the 
first row, % is plotted with 
$m=1\times10^{-22}\rm{eV}/c^2$, %the 
second row, % is plotted with 
$m=2\times10^{-22}\rm{eV}/c^2$ and %the 
third row, % is plotted with 
$m=5\times10^{-22}\rm{eV}/c^2$; %the 
first column gives %is the 
normalized density profile while %and 
the second column represents %is 
the physical density profile; %the 
black curves mark pure FDM solitons %is plotted 
without the background baryon profile, %the 
red dotted curves show the effect of %is plotted with 
$\tilde{\rho}_{\rm bg}^x$, %the 
blue dashed curves, that of % is plotted with 
$\tilde{\rho}_{\rm bg}^y$ and %the 
green dash-dotted curves, of % is plotted with 
$\tilde{\rho}_{\rm bg}^z$.}
\label{fig:prof2}
\end{figure*}

In Tabs.~\ref{tb:bx1}-\ref{tb:bz2}, we list the eigenvalues $\tilde{\gamma}$, ground state solution %the 
soliton mass $\tilde{M}$ and %of the ground state solutions and %the 
final value %s %of 
$\lambda_0$ for %the different background baryon profile $\tilde{\rho}_{\rm bg}^i$ and %the 
each FDM mass %of FDM 
$m$, background baryon profile $\tilde{\rho}_{\rm bg}^i$ and soliton physical mass $M$, respectively.
We %Meanwhile, we 
plot the normalized and the physical density profiles of the ground state solutions with different background baryon profile $\tilde{\rho}_{\rm bg}^i$ for various %and the mass of 
FDM masses $m$ and soliton physical masses $M$ in Fig.~\ref{fig:prof1} and Fig.~\ref{fig:prof2}, respectively. Comparing the soliton profiles corresponding to each %with different 
$\tilde{\rho}_{\rm bg}^i$ %plotted 
in every subplot, we find that the fine-structure of the background baryon profile does not radically change the soliton profiles and that %matter and 
our treatment for the ellipsoidal baryon profile is reasonable. Contrasting %And again, comparing 
the soliton profiles with different $m$, we find that %the 
heavier %\Mov{the }
FDM particles %, \Mov{the denser, }%
form a 
denser, more compact %, %denser 
and insensitive to the background baryon profile, %the 
soliton%\Mov{ that they form becomes}%, which is more insensitive to the background baryon profile
.
Since, for the Milky Way, the soliton physical masses $M$ derived from two soliton-halo mass relations Eqs.~\eqref{eq:r_h} and \eqref{eq:r_h_n} are very similar, our results do not obviously depend on the choice of the soliton-halo mass relation.

\begin{table}[htbp] 
\renewcommand\arraystretch{1.5}
\captionsetup{justification=raggedright}
\caption{Eigenvalue %Given the background baryon profile $\tilde{\rho}_{\rm bg}^x$ and the mass of FDM $m$, the corresponding eigenvalue 
$\tilde{\gamma}$, ground state solution % and 
soliton mass $\tilde{M}$ and %of the ground state solutions as well as the 
final values %of 
$\lambda_0$ for a set of FDM masses $m$, in the case of the background baryon profile $\tilde{\rho}_{\rm bg}^x$ and using the soliton-halo mass relation Eq.~\eqref{eq:r_h}.} 
\label{tb:bx1}
\begin{tabular}{ | c | c | c | c | } 
\hline 
$m$ & $\tilde{\gamma}$ & $\tilde{M}$& $\lambda_{0}$\\ 
\hline  
$1\times10^{-22}\rm{eV}/c^2$& 0.7600 & 1.4807& $4.3088\times10^{-7}$\\ 
$2\times 10^{-22}\rm{eV}/c^2$& 0.7037 & 1.7103& $3.2300\times10^{-7}$\\ 
$5\times 10^{-22}\rm{eV}/c^2$& 0.6644 & 1.9471& $2.4920\times10^{-7}$\\       
\hline
\end{tabular}
\end{table}

\begin{table}[htbp] 
\renewcommand\arraystretch{1.5}
\captionsetup{justification=raggedright}
\caption{Eigenvalue %Given the background baryon profile $\tilde{\rho}_{\rm bg}^y$ and the mass of FDM $m$, the corresponding eigenvalue 
$\tilde{\gamma}$, ground state solution % and 
soliton mass $\tilde{M}$ and %of the ground state solutions as well as the 
final values %of 
$\lambda_0$ for a set of FDM masses $m$, in the case of the background baryon profile $\tilde{\rho}_{\rm bg}^y$ and using the soliton-halo mass relation Eq.~\eqref{eq:r_h}.} 
\label{tb:by1}
\begin{tabular}{ | c | c | c | c | } 
\hline $m$ & $\tilde{\gamma}$ & $\tilde{M}$& $\lambda_{0}$\\ 
\hline  
$1\times10^{-22}\rm{eV}/c^2$& 0.7521 & 1.6009& $3.6863\times10^{-7}$\\ 
$2\times 10^{-22}\rm{eV}/c^2$& 0.7022 & 1.7670& $3.0256\times10^{-7}$\\ 
$5\times 10^{-22}\rm{eV}/c^2$& 0.6642 & 1.9483& $2.4889\times10^{-7}$\\       
\hline
\end{tabular}
\end{table}

\begin{table}[htbp] 
\renewcommand\arraystretch{1.5}
\captionsetup{justification=raggedright}
\caption{Eigenvalue %Given the background baryon profile $\tilde{\rho}_{\rm bg}^z$ and the mass of FDM $m$, the corresponding eigenvalue 
$\tilde{\gamma}$, ground state solution % and 
soliton mass $\tilde{M}$ and %of the ground state solutions as well as the 
final values %of 
$\lambda_0$ for a set of FDM masses $m$, in the case of the background baryon profile $\tilde{\rho}_{\rm bg}^z$ and using the soliton-halo mass relation Eq.~\eqref{eq:r_h}.} 
\label{tb:bz1}
\begin{tabular}{ | c | c | c | c | } 
\hline $m$ & $\tilde{\gamma}$ & $\tilde{M}$& $\lambda_{0}$\\ 
\hline  
$1\times10^{-22}\rm{eV}/c^2$& 0.7265 & 1.6652& $3.4068\times10^{-7}$\\ 
$2\times 10^{-22}\rm{eV}/c^2$& 0.6859 & 1.8400& $2.7970\times10^{-7}$\\ 
$5\times 10^{-22}\rm{eV}/c^2$& 0.6596 & 1.9922& $2.3805\times10^{-7}$\\       
\hline
\end{tabular}
\end{table}

\begin{table}[htbp] 
\renewcommand\arraystretch{1.5}
\captionsetup{justification=raggedright}
\caption{Eigenvalue %Given the background baryon profile $\tilde{\rho}_{\rm bg}^x$ and the mass of FDM $m$, the corresponding eigenvalue 
$\tilde{\gamma}$, ground state solution % and 
soliton mass $\tilde{M}$ and %of the ground state solutions as well as the 
final values %of 
$\lambda_0$ for a set of FDM masses $m$, in the case of the background baryon profile $\tilde{\rho}_{\rm bg}^x$ and using the soliton-halo mass relation Eq.~\eqref{eq:r_h_n}.} 
\label{tb:bx2}
\begin{tabular}{ | c | c | c | c | } 
\hline 
$m$ & $\tilde{\gamma}$ & $\tilde{M}$& $\lambda_{0}$\\ 
\hline  
$1\times10^{-22}\rm{eV}/c^2$& 0.7773 & 1.4228& $3.9834\times10^{-7}$\\ 
$2\times 10^{-22}\rm{eV}/c^2$& 0.6922 & 1.7784& $3.7065\times10^{-7}$\\ 
$5\times 10^{-22}\rm{eV}/c^2$& 0.6537 & 2.0275& $4.6674\times10^{-7}$\\       
\hline
\end{tabular}
\end{table}

\begin{table}[htbp] 
\renewcommand\arraystretch{1.5}
\captionsetup{justification=raggedright}
\caption{Eigenvalue %Given the background baryon profile $\tilde{\rho}_{\rm bg}^y$ and the mass of FDM $m$, the corresponding eigenvalue 
$\tilde{\gamma}$, ground state solution % and 
soliton mass $\tilde{M}$ and %of the ground state solutions as well as the 
final values %of 
$\lambda_0$ for a set of FDM masses $m$, in the case of the background baryon profile $\tilde{\rho}_{\rm bg}^y$ and using the soliton-halo mass relation Eq.~\eqref{eq:r_h_n}.} 
\label{tb:by2}
\begin{tabular}{ | c | c | c | c | } 
\hline $m$ & $\tilde{\gamma}$ & $\tilde{M}$& $\lambda_{0}$\\ 
\hline  
$1\times10^{-22}\rm{eV}/c^2$& 0.7682 & 1.5464& $3.3721\times10^{-7}$\\ 
$2\times 10^{-22}\rm{eV}/c^2$& 0.6898 & 1.8245& $3.5218\times10^{-7}$\\ 
$5\times 10^{-22}\rm{eV}/c^2$& 0.6531 &  2.0265& $4.6720\times10^{-7}$\\       
\hline
\end{tabular}
\end{table}

\begin{table}[htbp] 
\renewcommand\arraystretch{1.5}
\captionsetup{justification=raggedright}
\caption{Eigenvalue %Given the background baryon profile $\tilde{\rho}_{\rm bg}^z$ and the mass of FDM $m$, the corresponding eigenvalue 
$\tilde{\gamma}$, ground state solution % and 
soliton mass $\tilde{M}$ and %of the ground state solutions as well as the 
final values %of 
$\lambda_0$ for a set of FDM masses $m$, in the case of the background baryon profile $\tilde{\rho}_{\rm bg}^z$ and using the soliton-halo mass relation Eq.~\eqref{eq:r_h_n}.} 
\label{tb:bz2}
\begin{tabular}{ | c | c | c | c | } 
\hline $m$ & $\tilde{\gamma}$ & $\tilde{M}$& $\lambda_{0}$\\ 
\hline  
$1\times10^{-22}\rm{eV}/c^2$& 0.7395 & 1.6119& $3.1036\times10^{-7}$\\ 
$2\times 10^{-22}\rm{eV}/c^2$& 0.6767 & 1.8899& $3.2823\times10^{-7}$\\ 
$5\times 10^{-22}\rm{eV}/c^2$& 0.6523 & 2.0411& $4.6054\times10^{-7}$\\       
\hline
\end{tabular}
\end{table}

\section{Summary and Discussion}
\label{sec:sd}
In this paper, we present %enumerate some 
reasonable variants of the exact SP system that predict FDM solitons in a non relativistic gravitational potential. Each variant aims at modeling a different effect, such as the variants due to a central supermassive BH (section~\ref{sec:ge}), the system's own angular momentum (section~\ref{sec:gm}), an extra denser and compact soliton (section~\ref{sec:edr}) and an ellipsoidal baryon background (section~\ref{sec:bar}). We approximated each of them by %All of them can be considered as 
an almost spherical system and solved them with the shooting method. We reach the same conclusion as~\cite{Davies:2019wgi}, that the %larger 
central supermassive BH renders %makes 
the soliton %more 
denser and more compact. For the first time, we %prove 
demonstrate that the effect of the gravitomagnetic field derived from the soliton's self-angular momentum can be neglected compared to the contributions from the 
soliton's self-gravitating %gravitoelectric 
field, while the effect of gravitomagnetic field derived from the central supermassive BH angular momentum %, on the contrary, 
is comparable to the contribution from the gravitoelectric field from that BH. %Incidentally, it corresponds 
% which is the only 
%\Mov{the surviving gravoelectromagnetic effect that subsists in the purely }%survival in the 
%Newtonian regime. 
Unlike the interference or collision of two similar solitons, certain snapshots of the interaction between two solitons with an extreme density-ratio is relatively simple: the smaller soliton with higher density is almost un%not 
affected but the larger soliton with lower density shrinks dramatically. Finally, we revisit the topic discussed in~\cite{Bar:2019bqz}, showing that the baryon background is important for the soliton profile. We %But we 
propose a reasonable treatment for the ellipsoidal baryon background, from which we are able to % and then 
highlight the importance of the baryon background for lighter FDM particle.

Of course, we could %can 
enlarge the parameter space of the underlying theoretical model and produce a greater diversity in the soliton solution% of soliton
. For example, the self-interactions between FDM particles~\cite{Chavanis:2019bnu} or the modified gravitational potential due to the modified gravity~\cite{Chen:2024pyr} would result in a different set of coupled equations, hence a different soliton profile. In any case, the %Although the Universe has picked out the only underlying rule, these 
variants of the exact SP system presented in this paper are taking into account environmental effects, so the mapping between a basic system and its variants should be common to %the 
theoretical models.

\begin{acknowledgments}
%Ke Wang is supported by grants from NSFC (grant No.12247101). 
MLeD acknowledges the financial support by the Lanzhou University starting
fund, the Fundamental Research Funds for the Central Universities
(Grant No. lzujbky-2019-25) and NSFC (grant No.12247101).
\end{acknowledgments}

\end{document}